\documentclass[10pt,a4paper]{article}
\usepackage[utf8]{inputenc}
\usepackage[english]{babel}
\usepackage[affil-sl]{authblk}
\usepackage{amsmath,amssymb,mathrsfs}
\usepackage{hyperref}
\usepackage{amsfonts}
\usepackage{makeidx}
\usepackage{multicol}
\usepackage{balance}
\usepackage{appendix}
\usepackage[left=2cm,right=2cm,top=2cm,bottom=2cm]{geometry}
\usepackage{abstract}
\usepackage{graphicx}
\usepackage{float}
\usepackage{lmodern}
\usepackage{fourier}
\usepackage{cite}
\usepackage{abstract}
\usepackage{booktabs}
\begin{document}
\title{Fourier optics: basic concepts}
\author{Stephane Perrin \thanks{Corresponding author: \href{mailto:stephane.perrin@unistra.fr}{stephane.perrin@unistra.fr}}~}
\author{Paul Montgomery}
\affil{ICube, UMR CNRS 7357 - University of Strasbourg, FR-67412 ILLKIRCH}
\maketitle
\balance
\textbf{Based on diffraction theory and the propagation of the light, Fourier optics is a powerful tool allowing the estimation of a visible-range imaging system to transfer the spatial frequency components of an object. The analyses of the imaging systems can thus be performed and the the performance retrieved. For a better understanding of the optical study, I present a short introduction of the Fourier optics and I review the mathematical treatment depending on the illumination conditions of the imaging system. Furthermore, resolution criteria based on Fourier optics are detailed. Also, the impact of aberrations on the imaging quality are discussed.}
\begin{multicols}{2}
\subsection*{INTRODUCTION}
Performance of a linear optical system can be evaluated using Fourier optics \cite{Malacara03}. Indeed, the ability of a single lens or a more complex system to reproduce an image of an (1D or 2D) object can be quantify by decomposing the object in Fourier series. In Fourier optics, the object is thus considered as a sum of spatial sinusoidal waves at specific frequencies by analogy with electronics \cite{Goodman96}. Despite the resolution limit of a microscope by the diffraction of the light was explicitly mentioned by E. Verdet in 1869 \cite{Verdet69}, the concept of Fourier series decomposition appeared firstly in 1873 with the works of E. Abbe \cite{Abbe74}. He considered the objects as periodic diffraction elements and he showed that at least two orders of a diffraction grating should be collected by the objective lens of a microscope in order to barely resolve the features of the grating. He describes in words the first definition of the spatial resolution $\delta_{x,y}$ which can be mathematically expressed as:
\begin{equation}
\delta_{x,y}=\frac{\lambda}{n~\sin\left(\theta\right)}=\frac{\lambda}{NA}
\label{EqAbbe1}
\end{equation}
Where $\lambda$ is the wavelength of the monochromatic light source and $\theta$, the diffraction angle of the periodic object. According to the Abbe theory, only diffracted components generated by the object having a spatial period higher than $\delta_{x,y}$ are intercepted by the finite pupil of the lens. Thus, the features with higher spatial frequencies are not resolved. In addition, the term of numerical aperture (NA) were firstly introduced as the sine of the diffraction angle $\theta$ \cite{Abbe81}. Moreover, E. Abbe reported that the resolution $\delta_{x,y}$ can be improved using an focusing (or oblique) illumination of the object. This assumption allows determining the total NA of a microscope as the sum of the NAs of each optical component of the microscope. If the NA of the illumination part $NA_{i}$ equals the NA of the collection part $NA_{c}$, Eq.~\ref{EqAbbe1} can hence be written: 
\begin{equation}
\delta_{x,y}=\frac{\lambda}{NA_{i} + NA_{c}}=\frac{\lambda}{2~NA}
\label{EqAbbe2}
\end{equation}
It should be noted that a direct focusing illumination through a Abbe condenser lens has however the disadvantage of imaging the lamp filament on the sample. A. Köhler took an interest in illumination conditions for microscopy and developed thus a method for homogenising the intensity of the incident beam using an arrangement of lenses and diaphragms \cite{Kohler93}.\\ 
\indent In 1876, H. von Helmholtz confirmed the Abbe theory through a mathematical demonstration of the resolution limit \cite{Helmholtz76}. At the same time, he exposed the impact of the coherence of the light for avoiding the phase relations \cite{Lauterbach12} and he showed that Eq.~\ref{EqAbbe1} also requires a factor $1/2$ when the illumination light source is incoherent ($\lambda$ becomes thus $\lambda_{0}$, the central wavelength of the broadband light source). Furthermore, H. von Helmholtz discussed a new definition of the spatial resolution, reported previously in 1874 by Lord Rayleigh \cite{Rayleigh74}. \\
\indent J. W. Rayleigh did not treat microscopic objects as grating elements illuminated with plane waves but rather as as a sum of white-light-emitting point sources. Using the Airy works who calculated the diffraction pattern of a bright point source \cite{Airy35}, J. W. Rayleigh defined the angular separation of an imaging system as the distance between the intensity maximum and the first intensity minimum of the diffraction pattern of an on-axis point source\footnote{The diffraction image of an emitting incoherent point source through an optical system is the point spread function (PSF) of the system.}. Based on Fourier’s theorem, this definition leads to a new resolution expression:
\begin{equation}
\delta_{x,y}\approx0.61\frac{f~\lambda}{R}
\label{Rayleigh}
\end{equation}
Where $f$ is the focal length of the lens and $R$ is the radius of the finite pupil of the lens. Some years later, in 1879, Lord Rayleigh presented a view equivalent to regarding diffraction effects as resulting from the exit pupil \cite{Rayleigh79}. In 1896, Lord Rayleigh extended his investigations to different objects (points, lines and gratings) and aperture shapes \cite{Rayleigh96}. \\
\indent Twenty years later, based on the limits of photodetectors, C. M. Sparrow defined the smallest recognizable inter-space as the distance for which the irradiance pattern of two incoherent point sources has no curvature in the center \cite{Sparrow16}, giving the expression:
\begin{equation}
\delta_{x,y}=0.47\frac{\lambda}{NA}
\label{Sparrow}
\end{equation}
These criteria assume an aberrations-free imaging system. Nevertheless, in practice, evaluating the resolution is more complex due to the noise and the aberrations of optical components. Furthermore, the nature of the light source, i.e. coherent, partially coherent, partially incoherent or incoherent, should be considered. In 1927, W. V. Houston proposes thus to use the full-width at half-maximum (FWHM) of the diffraction pattern of a point source (being the PSF of the imaging system), to quantify the lateral resolution \cite{Houston27} because more useful in practice and also applicable to diffraction patterns that do not fall off to zero, i.e. a Gaussian PSF or a Lorentzian PSF. \\
\indent In 1946, P.M. Duffieux introduced Fourier optics for evaluating the spatial frequency transfers through the optical system using sinusoidal test patterns \cite{Duffieux46}. And, latter, H.H. Hopkins led the way in the use of transfer function methods for the assessment of the quality of optical imaging systems, making the analogy with analog electronic systems. In 1960, V. Ronchi highlighted the importance of considering both the sensitivity of the sensor and the illumination conditions, to determine the resolving power of an imaging system \cite{Ronchi61}.\\
\indent Nowadays, Fourier optics is often used for the design of new optical components or the analysis of imaging system. And the mathematical treatment of Fourier optics was demonstrated and is supposed to be known \cite{Saleh91}. However, the nature of the light dependence is often neglected and not considered in the literature. Thus, this manuscript reviews the influence of the coherence of the light source on the transfer function of a visible-range (or infra-red) imaging system. The resolution criteria based on Fourier optics are explained. Furthermore, the effects of aberrations on the imaging quality are discussed.
\subsection*{IMAGING SYSTEMS}
\begin{figure}[H]
\centering
\includegraphics[width=8cm]{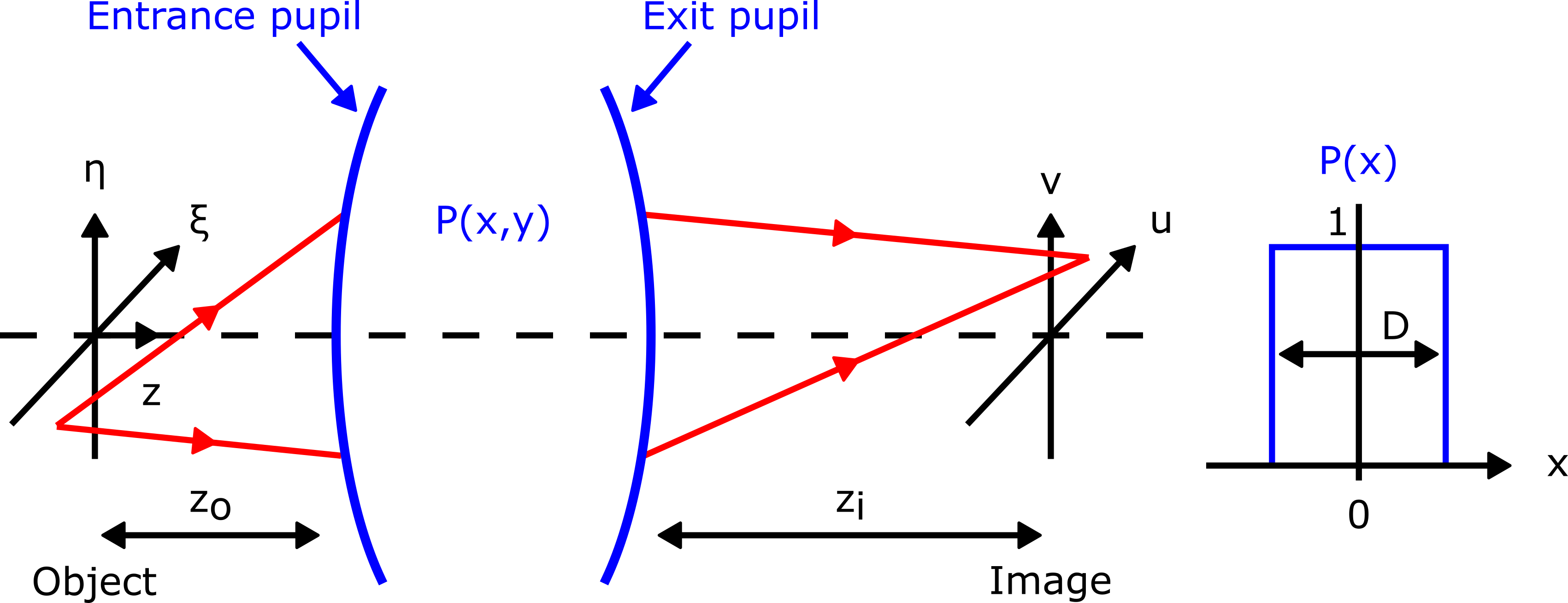}
\caption{\textbf{Model of an imaging system and representation of its pupil function.} The electric field from the emitting point source in the object plane is transmitted by the imaging system in the image plane. The imaging system has a normalized pupil function $P(x,y)$ with a diameter $D$. $\xi$ and $\eta$ are the Cartesian coordinates in the object plane. $x$ and $y$ are the Cartesian coordinates in the principal planes of the imaging system. $u$ and $v$ are the Cartesian coordinates in the image plane. $z$ is the optical axis.}
\label{model}
\end{figure}

In Fourier optics, the imaging systems are supposed time invariant and linear. Figure. \ref{model} represents a generalized scheme of an optical system, i.e. an assembly of optical elements, collecting the electric field from a point source placed in the object plane and propagating it in the image plane. The object plane is placed at a distance $z_{o}$ of the entrance pupil, i.e. the working distance, and the imaging plane at a distance $z_{i}$ of the exit pupil. The exit pupil (or simply the finite circular pupil function $P(x,y)$) of the imaging system is considered in the following diffraction equations. Indeed, only the output pupil plays a role in the diffraction of the light. The pupil function $P(x,y)$ has a diameter $D$ and is unity inside and zero outside the projected aperture. Gauss approximations are considered, meaning that the incident angles are small and the point light sources are close to the optical axis\footnote{Formerly, a Gaussian image was an image formed from these approximations}.
Furthermore, the imaging system is first assumed to be free of aberration, i.e. a diffraction-limited optical system. An imaging system is said to be diffraction-limited if a diverging spherical wave, emanating from a point-source object, is converted by the system into a new wave, again perfectly spherical, that converges towards an ideal point in the image plane, where the location of that ideal image point is related to the location of the original object point through a simple scaling factor, i.e. the magnification $M$. The magnification factor must be the same for all points in the lateral field of view of the imaging system \cite{Goodman96}. \\
\indent In the image plane, the 2D complex amplitude distribution of the image of a point light source is represented by a superposition integral,
\begin{equation}
U_{i}\left(u,v\right) = \displaystyle \iint_{\infty}^{\infty} h\left(u,v;\xi,\eta\right) \times U_{o}\left(\xi,\eta\right) \, \mathrm{d}\xi \mathrm{d}\eta,
\label{eq1}
\end{equation}
with $U_{o}\left(\xi,\eta\right)$, the complex electric field of the emitting point source. The coordinates $u$ and $v$ are given by $u=M\xi$ and $v=M\eta$ where the magnification $M$ of the imaging system could be negative or positive. Using the convolution theorem \cite{conv}, Eq. \ref{eq1} can be rewritten:
\begin{equation}
U_{i}\left(u,v\right) = h\left(u,v;\xi,\eta\right) \circledast U_{o}\left(\xi,\eta\right)
\label{eq2}
\end{equation}
The amplitude response to a point-source object $h\left(u,v;\xi,\eta\right)$ of the imaging system (also called amplitude point spread function) at a position $\left(\xi,\eta\right)$ is defined as the Fourier transform of its pupil function $P(x,y)$.
\begin{align}
h\left(u,v;\xi,\eta\right) = \frac{A}{\lambda z_{i}} \iint_{\infty}^{\infty} P\left(x,y\right) \nonumber \times \\
\exp\left\lbrace -j \frac{2\pi}{\lambda z_{i}} \left[ \left(u-M\xi\right)x+\left(v-M\eta\right)y \right]\right\rbrace \, \mathrm{d}x \mathrm{d}y
\label{eq3}
\end{align}
In order to determine the irradiance $I_{i}\left(u,v\right)$ recorded by a photo-detector placed in the image plane, the square of the image amplitude is time-averaged.
\begin{equation}
I_{i}\left(u,v\right) = \left\langle |U_{i}\left(u,v\right)|^{2}\right\rangle
\label{eq4}
\end{equation}
\end{multicols}
\begin{figure*}[!h]
\centering
\includegraphics[width=12cm]{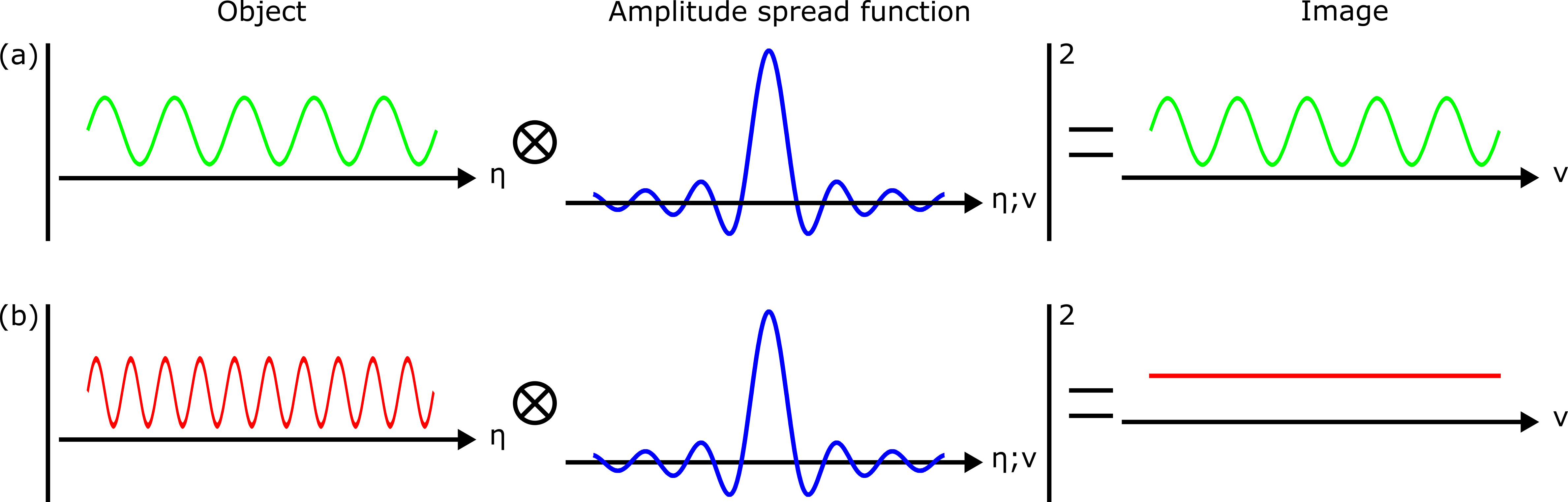}
\caption{\textbf{Spatial-domain response of an optical system in coherent imaging.} The frequency of (a) the green object signal is lower than the frequency of (b) the red object signal. The image intensity is the square of the convolution product of the amplitude object with the amplitude spread function of the system.}
\label{asf}
\end{figure*}
\begin{figure*}[!h]
\centering
\includegraphics[width=12cm]{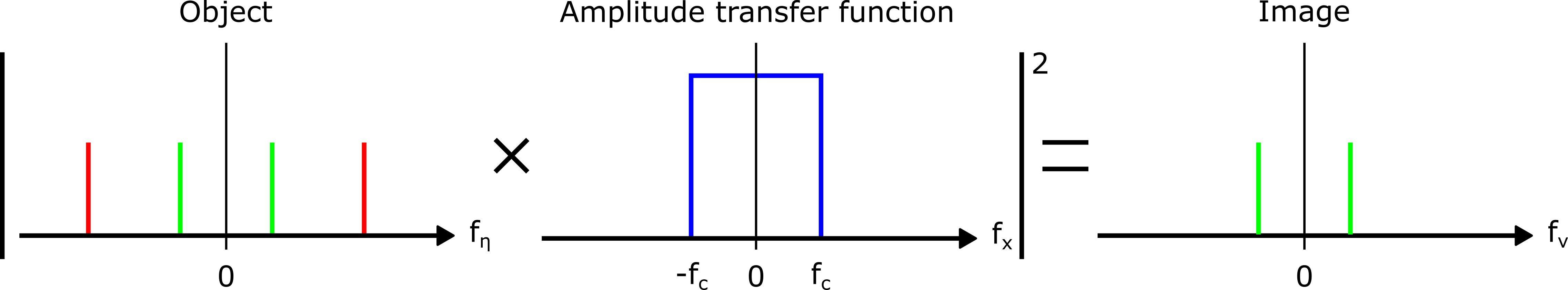}
\caption{\textbf{Frequency-domain response of an optical system in coherent imaging.} The frequency of the green object signal is lower than the frequency of the higher object signal. The image is the square of the product of the pupil function of the system with the amplitude object.}
\label{atf}
\end{figure*}
\begin{multicols}{2}
\subsection*{COHERENT IMAGING}
The coherent illumination of an object yields an imaging system linear in complex amplitude. Indeed, by defining a time-invariant phasor amplitude $U_{i}\left(u,v\right)$ in the image space, the imaging system is found to be described by an amplitude convolution equation. By combining the Eq.\ref{eq2} and Eq.\ref{eq4}, the expression of the image intensity can thus defined as the square of the convolution product of the object signal and the amplitude spread function $h\left(u,v;\xi,\eta\right)$. 
\begin{equation}
I_{i}\left(u,v\right) = \left\langle\big| h\left(u,v;\xi,\eta\right) \circledast U_{o}\left(\xi,\eta\right) \big|^{2} \right\rangle
\label{eq5}
\end{equation}
Figure \ref{asf} shows the image formation of an imaging system with coherent illumination in one dimensions. Assuming a rectangular-aperture amplitude distribution $P(x)$, the amplitude spread function $h\left(v;\eta\right)$, represented in blue, is a cardinal sinus function, i.e. a zero-order spherical Bessel function. Two objects (two continuous periodical waves) having a low and a high spatial frequencies are illustrated in green and in red, respectively. The imaging system transmits the low frequency object pattern keeping the same contrast (Fig.\ref{asf}(a)). Whereas, when the frequency of the entrance signal is higher (Fig.\ref{asf}(b)), the contrast of the image equals zero, i.e. the imaging system cannot resolve such frequency.
\indent Through a frequency analysis, a better visualization of the coherent imaging process is provided. Using convolution theorem and Eq. \ref{eq2}, the frequency-domain electric field in the image plane can be expressed by:
\begin{equation}
\mathscr{F} \left\lbrace U_{i}\left(u,v\right) \right\rbrace = \mathscr{F} \left\lbrace h\left(u,v;\xi,\eta\right) \right\rbrace \times \mathscr{F} \left\lbrace U_{o}\left(\xi,\eta\right) \right\rbrace,
\label{eq6}
\end{equation}
and, from Eq. \ref{eq3},
\begin{equation}
\mathscr{F} \left\lbrace U_{i}\left(u,v\right) \right\rbrace \propto P(x,y) \times \mathscr{F} \left\lbrace U_{o}\left(\xi,\eta\right) \right\rbrace,
\label{eq7}
\end{equation}
Then, the frequency-domain irradiance is defined by:
\begin{equation}
I_{i}\left(f_{y},f_{x}\right) \propto \big| P(x,y) \times \mathscr{F} \left\lbrace U_{o}\left(\xi,\eta\right) \right\rbrace\big|^{2}
\label{eq8}
\end{equation}
These equations show that the term of normalized amplitude transfer function $H\left(f_{x},f_{y}\right)$ of the imaging system is its pupil function $P(x,y)$.
\begin{align}
H\left(f_{x},f_{y}\right) &= \mathscr{F} \left\lbrace h\left(u,v;\xi,\eta\right) \right\rbrace \nonumber \\
&\propto \mathscr{F} \left\lbrace \mathscr{F} \left\lbrace P(x,y) \right\rbrace \right\rbrace \propto P \left( \lambda z_{i} f_{x}, \lambda z_{i} f_{y} \right)
\label{eq9}
\end{align}
Figure \ref{atf} illustrates the image formation in frequency domain of a finite-aperture imaging system using a coherent illumination. The object consists of two sinusoidal signals having a high and a low frequency equivalent to Fig. \ref{asf}.
When the frequency of the entrance object signal is below the cut-off frequency $f_{c}$ of the imaging system, e.g. green signal, the resulting contrast of its image stays unchanged. However, for higher frequency signal, e.g. red signal, the contrast drop to zero. The frequencies of the object being higher than the cut-off frequency of transfer function of the coherent imaging system are thus not resolved. The cut-off frequency $f_{c}$ of a perfectly-coherent imaging system is given by:
\begin{equation}
f_{c} = \frac{D}{\lambda z_{i}} = \frac{NA}{\lambda}
\label{eq10}
\end{equation}
This formula remind us the Abbe theory with Eq. \ref{EqAbbe1}. However, this assumption of strictly monochromatic illumination is overly restrictive. The illumination generated by real optical sources, including the LASER, are never perfectly monochromatic. The value of the cut-off frequency could hence be slightly reduced.
\end{multicols}
\begin{figure*}[!h]
\centering
\includegraphics[width=12cm]{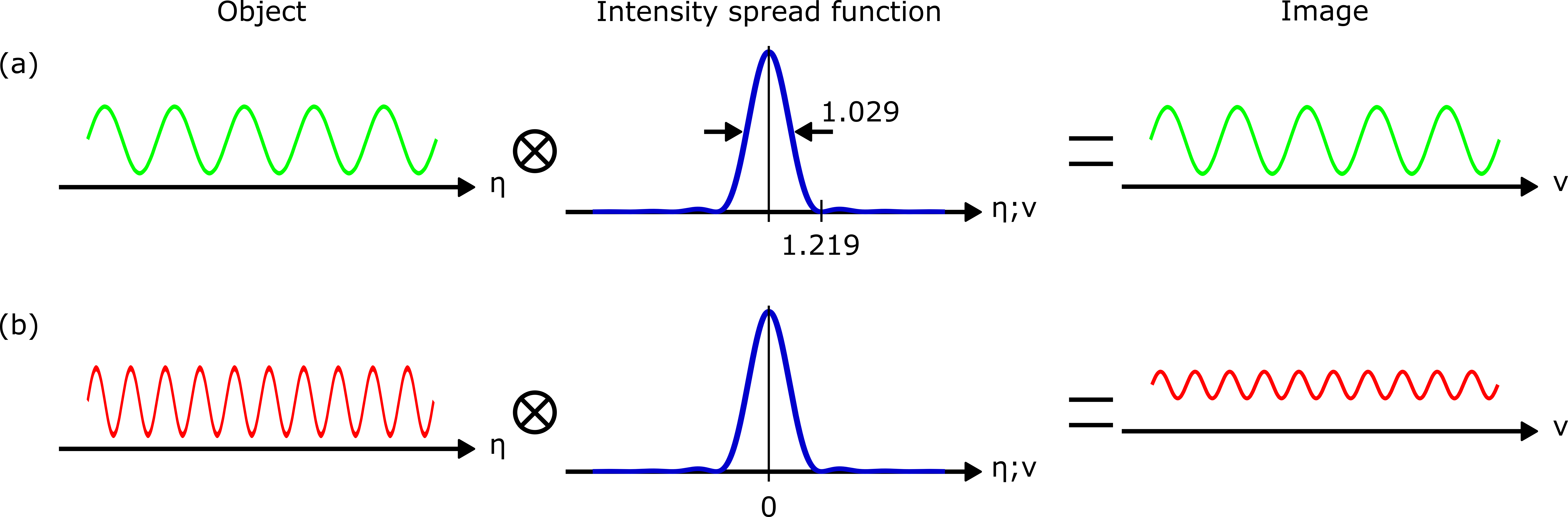}
\caption{\textbf{Spatial-domain response of an optical system in incoherent imaging.} The frequency of (a) the green object signal is lower than the frequency of (b) the red object signal. The intensity image is the convolution product of the intensity object with the intensity spread function of the system. For an aberration-free imaging system, the intensity spread function is an Airy disk where the FWHM equals 1.029 and the first zero is at 1.219 precisely.}
\label{isf}
\end{figure*}
\begin{figure*}[!h]
\centering
\includegraphics[width=12cm]{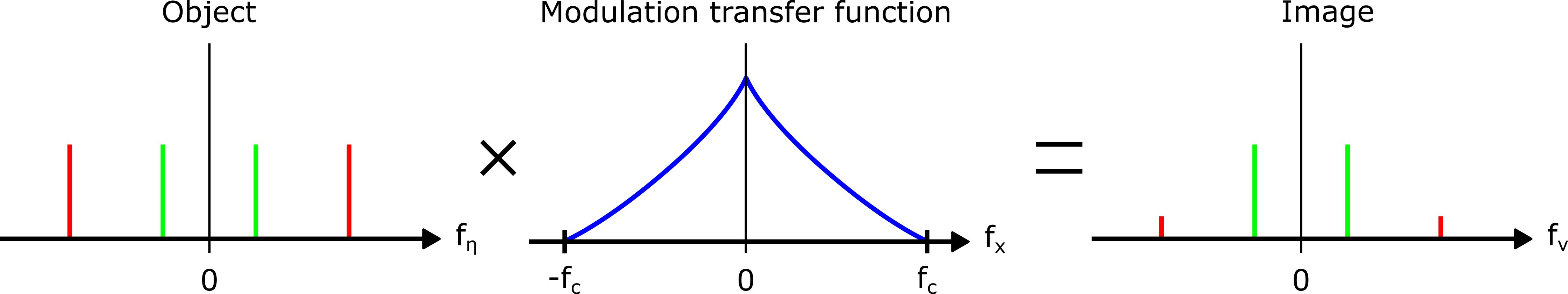}
\caption{\textbf{Frequency-domain response of an optical system in incoherent imaging.} The frequency of the green object signal is lower than the frequency of the higher object signal. The intensity image is the product of the modulation transfer function of the system with the square of amplitude object.}
\label{mtf}
\end{figure*}
\begin{multicols}{2}
\subsection*{INCOHERENT IMAGING}
Using an incoherent illumination, the image formation by the optical system is linear in intensity. The resulting intensity in the image plane is defined as the square of the amplitude spread function $h\left(u,v;\xi,\eta\right)$ convoluted at the object irradiance point source. Equation \ref{eq5} can thus be expressed:
\begin{equation}
I_{i}\left(u,v\right) = \big| h\left(u,v;\xi,\eta\right) \big|^{2} \circledast \big| U_{o}\left(\xi,\eta\right) \big|^{2}
\label{eq11}
\end{equation}
or,
\begin{equation}
I_{i}\left(u,v\right) = \big| h\left(u,v;\xi,\eta\right) \big|^{2} \circledast I_{o}\left(\xi,\eta\right)
\label{eq12}
\end{equation}
The imaging process of an aberration-free optical system using an incoherent light source is schemed in Fig.\ref{isf} in one dimension. The object having two distinct frequencies, as shown in Fig.\ref{asf}, are illustrated in green and in red. Here, the imaging system transmits the low-frequency object signal keeping the same contrast (Fig.\ref{isf}(a)). However, when the frequency of the entrance signal is higher (Fig.\ref{isf}(b)), the contrast of the image is decreased but still resolved by the imaging system. The incoherent spread function, also called intensity spread function or intensity point spread function (IPSF), is the square of the amplitude spread function $h\left(u,v;\xi,\eta\right)$ of the imaging system. This function is called Airy disk \cite{Airy35} where the full width at half maximum (FWHM) equals 1.029 and the first zero position is at $\eta = 1.219$. J. W. Rayleigh reported the use of the first zero of the Airy pattern in order to determine the distance $\delta_{x}$ between two resolved point sources \cite{Rayleigh74}.
\begin{equation}
\eta = 1.219 = \delta_{x}\frac{2~NA}{\lambda}
\end{equation}
In frequency domain, Eq. \ref{eq12} can be retrieved using the convolution theorem.
\begin{equation}
\mathscr{F} \left\lbrace I_{i}\left(u,v\right) \right\rbrace = \mathscr{F} \left\lbrace \big| h\left(u,v;\xi,\eta\right) \big|^{2} \right\rbrace \times \mathscr{F} \left\lbrace I_{o}\left(\xi,\eta\right) \right\rbrace,
\label{eq13}
\end{equation}
or
\begin{equation}
\mathscr{F} \left\lbrace I_{i}\left(u,v\right) \right\rbrace \propto \mathscr{F} \left\lbrace \big| \mathscr{F} \left\lbrace P(x,y) \right\rbrace \big|^{2} \right\rbrace \times \mathscr{F} \left\lbrace I_{o}\left(\xi,\eta\right) \right\rbrace
\label{eq14}
\end{equation}
In this equations, the term of optical transfer function $OTF\left(f_{x},f_{y}\right)$ is revealed, being the normalized autocorrelation function of the amplitude transfer function. 
\begin{align}
OTF\left(f_{x},f_{y}\right) &= \mathscr{F} \left\lbrace \big| h\left(u,v;\xi,\eta\right) \big|^{2} \right\rbrace \nonumber \\
&\propto \mathscr{F} \left\lbrace \big| \mathscr{F} \left\lbrace P(x,y) \right\rbrace \big|^{2} \right\rbrace
\label{eq15}
\end{align}
From the Fourier transform operation, a real and an imaginary parts result, leading to the modulation transfer function (MTF) and the phase transfer function (PTF).
\begin{equation}
MTF = \big| OTF \big| 
\label{eq16}
\end{equation}
\begin{equation}
PTF = angle \left\lbrace OTF \right\rbrace 
\label{eq17}
\end{equation}
The PTF is liable for the transversal shift of the image. Whereas, the MTF represents the contrast (or the visibility) distribution ratio between the image $M_{image}$ with the object $M_{object}$ at a given spatial frequency $f_{x}$,$f_{y}$.
\begin{equation}
MTF\left(f_{x},f_{y}\right) = \frac{M_{image}}{M_{object}}\left(f_{x},f_{y}\right)
\end{equation}
Where the contrast $M\left(f_{x},f_{y}\right)$ is related to the intensity modulation depth of the sinusoid signal.
\begin{equation}
M\left(f_{x},f_{y}\right) = \frac{I_{max} - I_{max}}{I_{max} + I_{max}}\left(f_{x},f_{y}\right)
\end{equation}
Figure \ref{mtf} illustrates the transfer function of an incoherent imaging system with a rectangular pupil function $P(x,y)$. The object signals are sinusoidal functions with different frequencies (see Fig. \ref{isf}). An entrance signal with a frequency below the cut-off frequency $f_{c}$ of the transfer function of the imaging system can be transmitted without reducing of the contrast. However, unlike in coherent imaging, the contrast of higher frequency signal than $f_{c}$ can be decreased, e.g. the red higher frequency signal, and still resolved. The cut-off frequency $f_{c}$ of the transfer function in incoherent imaging is defined by:
\begin{equation}
f_{c} = 2\frac{NA}{\lambda_{0}},
\label{eq18}
\end{equation}
confirming the results reported previously by H. von Helmholtz \cite{Helmholtz76}. The long wavelengths tend thus to decrease the cut-off frequency value. While, shorter wavelengths increase $f_{c}$. For polychromatic illumination, the MTF is determined by averaging the sum of $MTF\left(f_{x},f_{y}\right)$ of each wavelength, i.e. additioning the MTFs over the bandwidth of the light source \cite{telescope}. In experiments, the resulting polychromatic transfer function is assumed similar to the one at the central wavelength $\lambda_{0}$ of the spectrum. \\
\indent Mathematically, the MTF function of an aberration-free imaging system, having a circular pupil, can be expressed as:
\begin{equation}
MTF\left(f_{x},f_{y}\right) = \frac{2}{\pi}\left\lbrace\phi\left(f_{x},f_{y}\right) -\cos\left[\phi\left(f_{x},f_{y}\right)\right]\sin\left[\phi\left(f_{x},f_{y}\right)\right]\right\rbrace 
\end{equation}
where, in one dimension,
\begin{equation}
\cos\left[\phi\left(f_{x}\right)\right] = \frac{f_{x}}{f_{c}}~~~~\text{and}~~~~\cos\left[\phi\left(f_{y}\right)\right] = \frac{f_{y}}{f_{c}}
\end{equation}
\subsection*{INFLUENCE OF THE ABERRATIONS}
\begin{figure}[H]
\centering
\includegraphics[width=8cm]{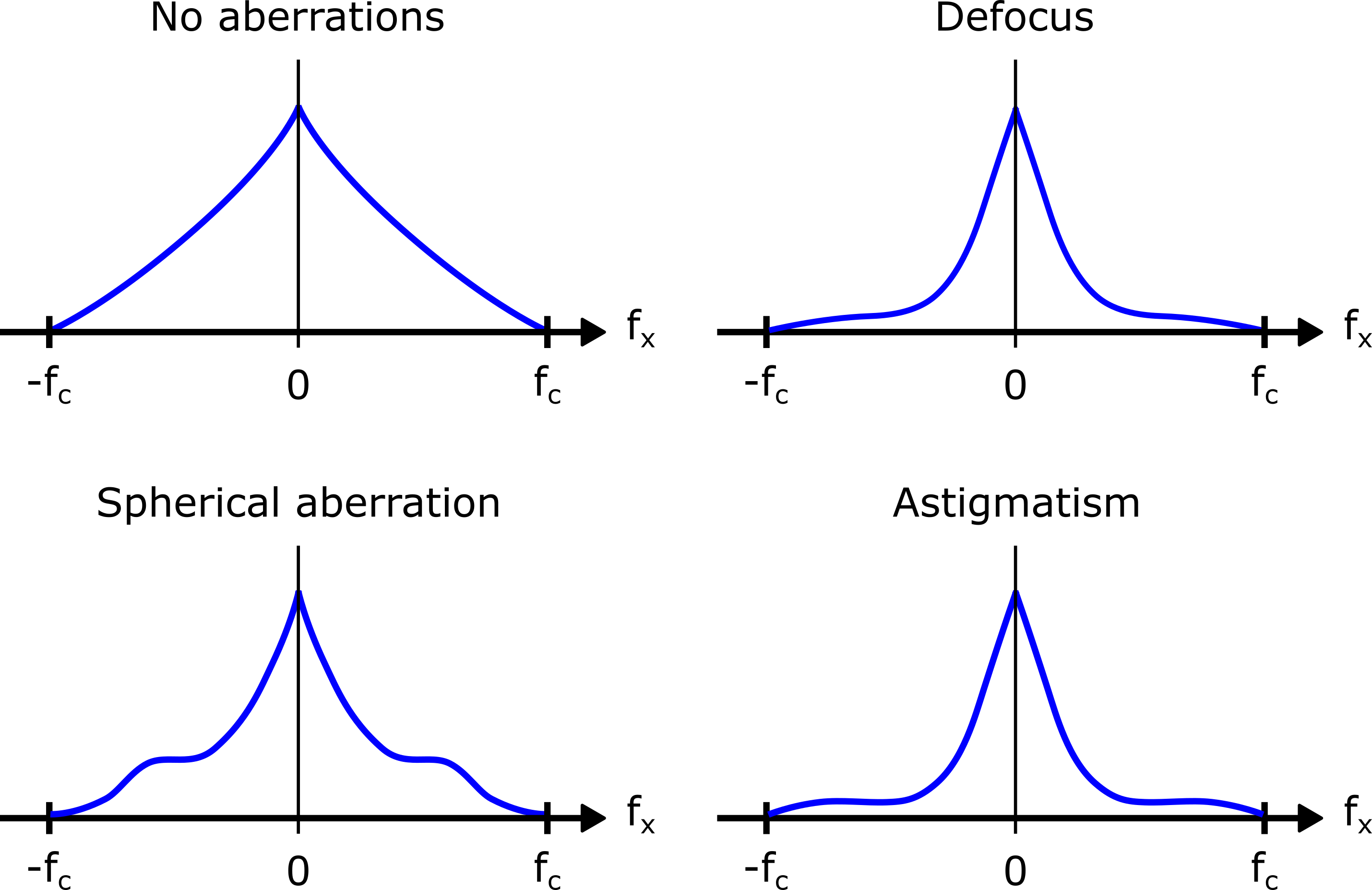}
\caption{\textbf{Modulation transfer function of an incoherent imaging system} being aberration-free or having defocus, spherical or astigmatism aberration.}
\label{aberration}
\end{figure}

Up to now, the manuscript relates the performance Fourier analysis of an aberration-free imaging system, i.e. a diffraction-limited system. In reality, the imaging systems provide often optical aberrations especially high-numerical-aperture systems. These aberrations decrease dramatically the performance of imaging systems through the deformation of the transfer function in both spatial and frequency domain. Indeed, in spacial domain, the point spread function can be widen, lowing the lateral resolution, or can be flattened, reducing the contrast of the image. Therefore, criteria exist for estimating the performance of an aberrated imaging system. \\
\indent The Strehl ratio (SR) allows a quantification of the contrast as a function of a perfect optical component, having the same numerical aperture \cite{Strehl95,Baranski14}, through a mathematical operation based on the ratio of intensity peaks.
\begin{equation}
SR = \frac{max \left(IPSF_{aberrated}\right)}{max \left(IPSF_{perfect}\right)}
\label{eq19}
\end{equation}
where $max \left(IPSF_{aberrated}\right)$ is the IPSF peak intensity of the system to be characterized and $max \left(IPSF_{perfect}\right)$ is the IPSF peak intensity of an aberration-free system. An system system is assumed diffraction limited only when the SR is higher than 0.8. Other criteria subsist and are listed in Tab. \ref{citerionPSF}.
\begin{table}[H]
\centering
\caption{\label{citerionPSF}Relation between SR and RMS of the wavefront.}
\begin{tabular}{c c c}
\toprule
{Strehl ratio} & RMS (wave) & Criterion \\
\midrule
0.96     & $\lambda / 32$ & $ $ \\
0.82     & $\lambda / 14$       & Maréchal \\
0.8      & $\lambda / 13.4$      & Diffraction limit  \\
\bottomrule
\end{tabular}
\end{table} 
According to Maréchal’s Strehl approximation, the root mean square (RMS) of the wavefront can then be deducted from SR when the aberrations are low (SR > 0.8).
\begin{equation}
SR = \exp\left[-\left(RMS_{Phi}\right)^{2}\right]
\label{eq20}
\end{equation}
In frequency domain, optical aberrations decrease the contrast of the MTF at some frequencies $f_{x,y}$ or yield to a linear lateral shift of the image pattern, i.e. PTF derivation. Figure \ref{aberration} illustrates the impact of the commonly optical aberrations on the MTF distribution. The cut-off frequency, depending on the wavelength of the light source and the NA of the imaging system, stays unchanged. However, the MTF of strongly-aberrated imaging system could fall off to zero at lower frequencies, leading perhaps, in practice, to wrong interpretation of the results.
H. H. Hopkins defined thus a criterion based on a drop-off contrast of the MTF equaling 20\%, by analogy to the Strehl ratio for IPSF \cite{telescope,Hopkins57}.
\subsection*{DISCUSSIONS}
First, periodic white and black bars, i.e. Ronchi rulings or USAF 1951 target, are widely preferred in experiments for example for the characterization of imaging system because more convenient to fabricate than sinusoidal grey-level pattern. The response of an alternating black and white lines is called contrast transfer function (CTF). By decomposing a one-dimension square-wave object in Fourier series, the CTF is related to the MTF by:
\begin{equation}
MTF\left(f_{x}\right) = \frac{\pi}{4} \sum_{n=0}^{N} \frac{1}{2n+1} CTF \left(\left(2n+1\right) f_{x}\right)
\end{equation}
However, the difference of the evolution curves between the MTF and the CTF is low \cite{telescope} and, in practice, is often not considered. \\
\indent Finally, in optical metrology, Fourier optics made it possible to measure the wavefront generated by an optical system using an iteration process of the propagation and the back-propagation of the light in several image planes \cite{Saxton72}. This characterization technique is called phase retrieval \cite{Fienup82,Perrin15}.
\subsection*{CONCLUSION}
This manuscript reviews the concept of Fourier optics for optical imaging. Therefore, the transfer function of finite-aperture imaging systems is detailed in both the spatial and the frequency domains considering the nature of the light. Indeed, the response of a coherent imaging system is linear in complex amplitude. Whereas, for incoherent imaging, the transfer function is related to the irradiance and the intensity. Also, the effect of optical aberrations on the imaging quality is discussed. Furthermore, resolution criteria are mathematically defined, assuming the sensor satisfies at least the Nyquist-based sampling. \\
\indent Recently, several imaging techniques have been developed in order to overcome the physical barrier of the diffraction of the light. Indeed, confocal microscopy [US patent \textbf{US3013467} (1961)], stimulated-emission-depletion fluorescence microscopy [Opt. Lett. \textbf{19}, 780 (1994)], scanning photonic jet scanning microscopy [Opt.
Express \textbf{15}, 17334 (2007)] or microsphere-assisted microscopy in 2D [Nat. Commun. \textbf{2}, 218 (2011)] and in 3D [Appl. Opt. \textbf{56}, 7249 (2017)] made the resolution possible to reach $\lambda$/7.

\end{multicols}
\newpage
\begin{appendices}
\textbf{Coherence effect on the irradiance}\\ \\
The coherence of the light can be divided in three cases: the spatial and the temporal coherence and the polarization. Here, only the two first cases are considered. The polarization of the light is neglected. An optical system behaves differently if illuminated by a temporally or spatially coherent or incoherent light. Indeed, a temporally incoherent illumination, typically a broadband light source, provides chromatic aberration which is typical evidence of temporal incoherence. The degree of spatial coherence alters the description of an optical system as a linear system. A coherent illumination is obtained whenever light appears to originate from a single point. The most common example of a source of such light is a laser, although more conventional sources, e.g. zirconium arc lamps, can yield coherent light if their output is first passed through a pinhole. Whereas, spatial incoherent light is obtained from diffuse or extended sources. We digress temporarily to consider the very important effects of polychromaticity. \\ \\
\indent In coherent imaging, the two wave sources have a constant initial phase difference and the same frequency. Thus, the intensity distribution $I$ resulting on the interference pattern from a wavefront division (temporally coherence) or an amplitude division (spatially coherence) can be written:
\begin{align}
I &= \big| a_{1}\exp \left( j\phi_{1} \right) + a_{2}\exp \left( j\phi_{2} \right) \big|^{2} \nonumber \\
&= I_{0} \left( 1+m \cos\left( \phi_{2} - \phi_{1} \right) \right)
\end{align}
with $j$ is the imaginary unit. The average intensity is given by $I_{0}=a_{1}^{2}+a_{1}^{2}=I_{1}+I_{2}$ and the contrast by $m=2\sqrt{I_{1}I_{2}}/\left(I_{1}+I_{2}\right)$. \\ \\
The phase delay $\Phi=\phi_{2}-\phi_{1}$ incurred in the paths of the two arms of the interferometer is: \\
 - $\phi_{2}-\phi_{1} = \frac{2\pi}{\lambda}\left(x_{2}-x_{1}\right)$ for an amplitude division interferometer,\\
 - $\phi_{2}-\phi_{1} = \frac{2\pi}{\lambda}\left( t_{2}-t_{1}\right)$ for a wavefront division interferometer.\\
$\lambda$ is the wavelength of the light source. \\ \\
\indent In incoherent imaging, the object illumination has the opposite property that the phasor amplitudes at all points on the object are varying in totally uncorrelated way. In this case, the phase delay $\Phi$ changes rapidly over the time ($\sum_{i} \Phi \left(\lambda_{i}\right) = 0$), leading to an intensity averaging out to zero:
\begin{equation}
I = \left\langle I_{0}\left( 1+m \cos\left( \phi_{2} - \phi_{1} \right) \right) \right\rangle
\end{equation}
\begin{equation}
I = I_{0}\left( 1+m \left\langle \cos\left( \phi_{2} - \phi_{1} \right) \right\rangle \right)
\end{equation}
\begin{equation}
I = I_{0} = a_{1}^{2}+a_{1}^{2}
\end{equation} \\
\indent However, in the perfectly coherent case, the intensity is computed as the modulus-squared of the sum of the phasors of the electric fields. The image intensity is thus linear in amplitude. Whereas, in the perfectly incoherent case, the intensity is computed as the sum of the modulus-squared of the phasors of the electric fields, leading to an intensity linearity of the image distribution. \\ \\
\indent In conclusion, if the illumination is coherent, the output field is described as the convolution of the input field with the amplitude spread function. While, if the illumination is incoherent, the output intensity is described as the convolution of the input intensity with the intensity spread function. 
\newpage
\textbf{Two dimensional representations of image formation using Fourier optics}.\\ \\
\begin{figure}[!h]
\centering
\includegraphics[scale=0.4]{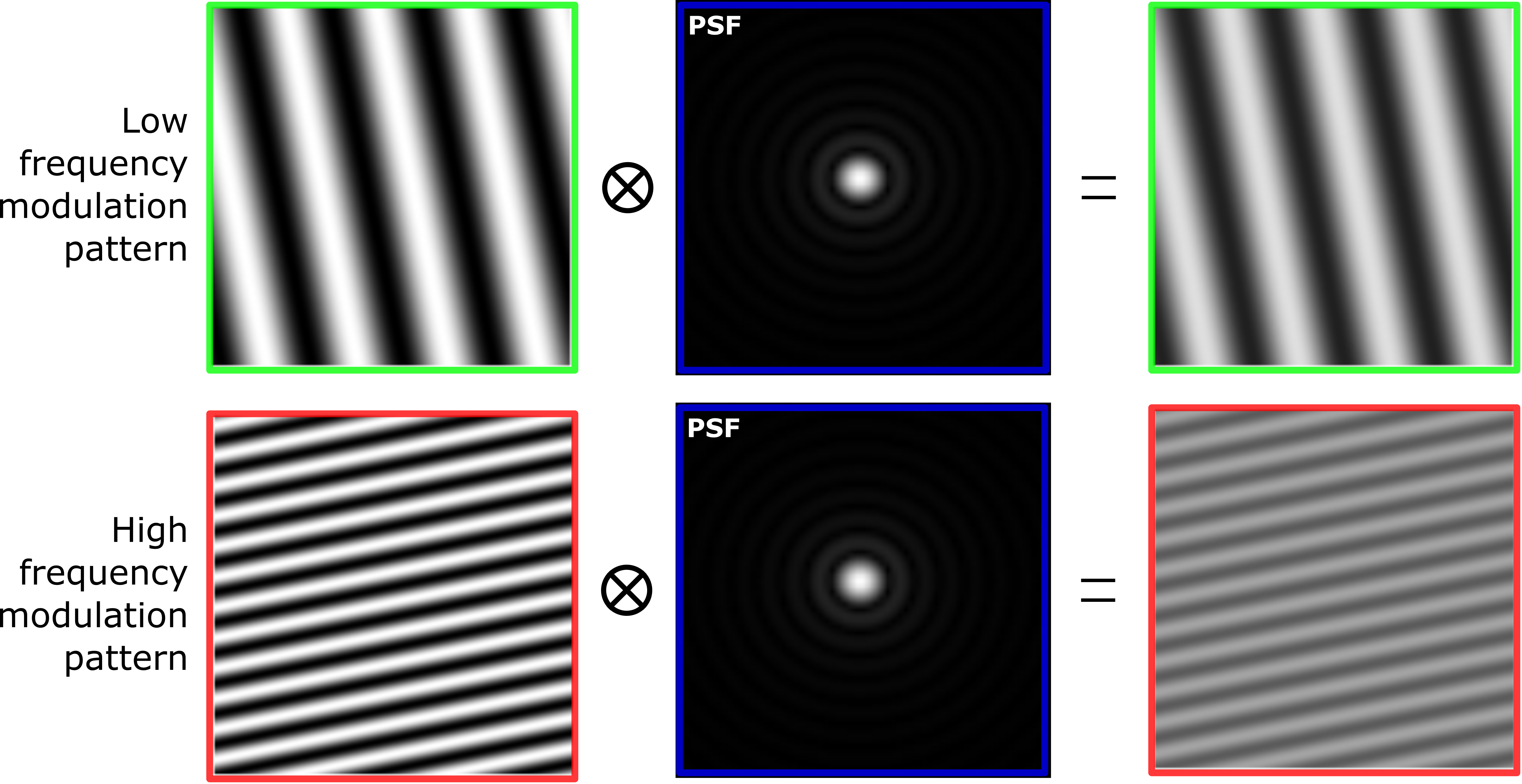}
\caption{Image formation of a low (in green) and a high (in red) frequency modulation pattern using the concept of Fourier optics in spatial domain. The illumination is incoherent and the imaging system, aberration-free. The imaging system resolves the two sinusoidal object patterns. However, the contrast of the resulting image of the higher-frequency pattern is reduced.}
\label{2dfrequency}
\end{figure}
\begin{figure}[!h]
\centering
\includegraphics[scale=0.4]{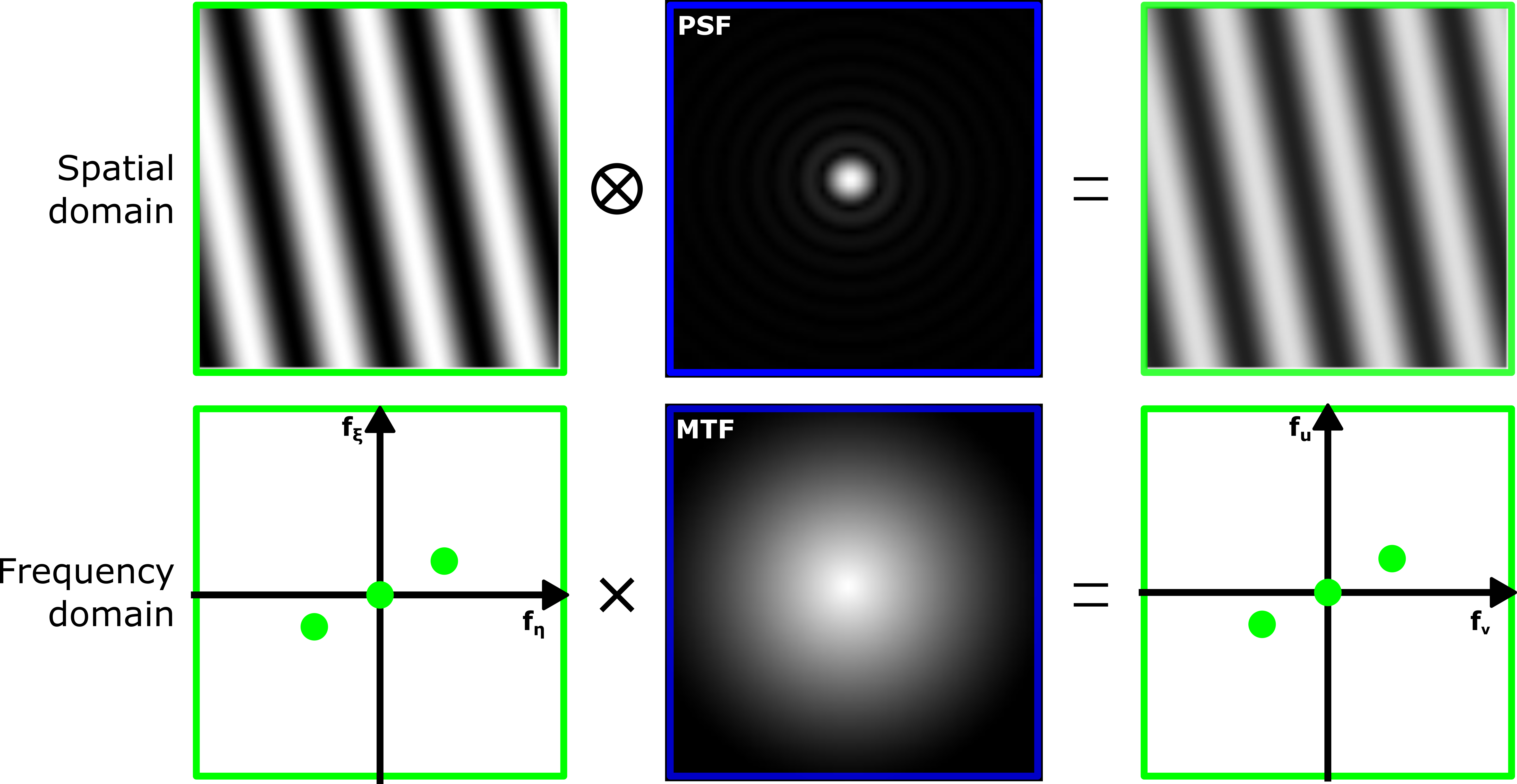}
\caption{Image formation of a low frequency modulation pattern using the concept of Fourier optics in spatial domain and in frequency domain. The illumination is incoherent and the imaging system, aberration-free. The imaging system resolves the sinusoidal object pattern. However, the contrast of the resulting image is reduced by the point spread function (in spatial domain) or the modulation transfer function (in frequency domain).}
\label{2ddomain}
\end{figure}
\begin{figure}[!h]
\centering
\includegraphics[scale=0.4]{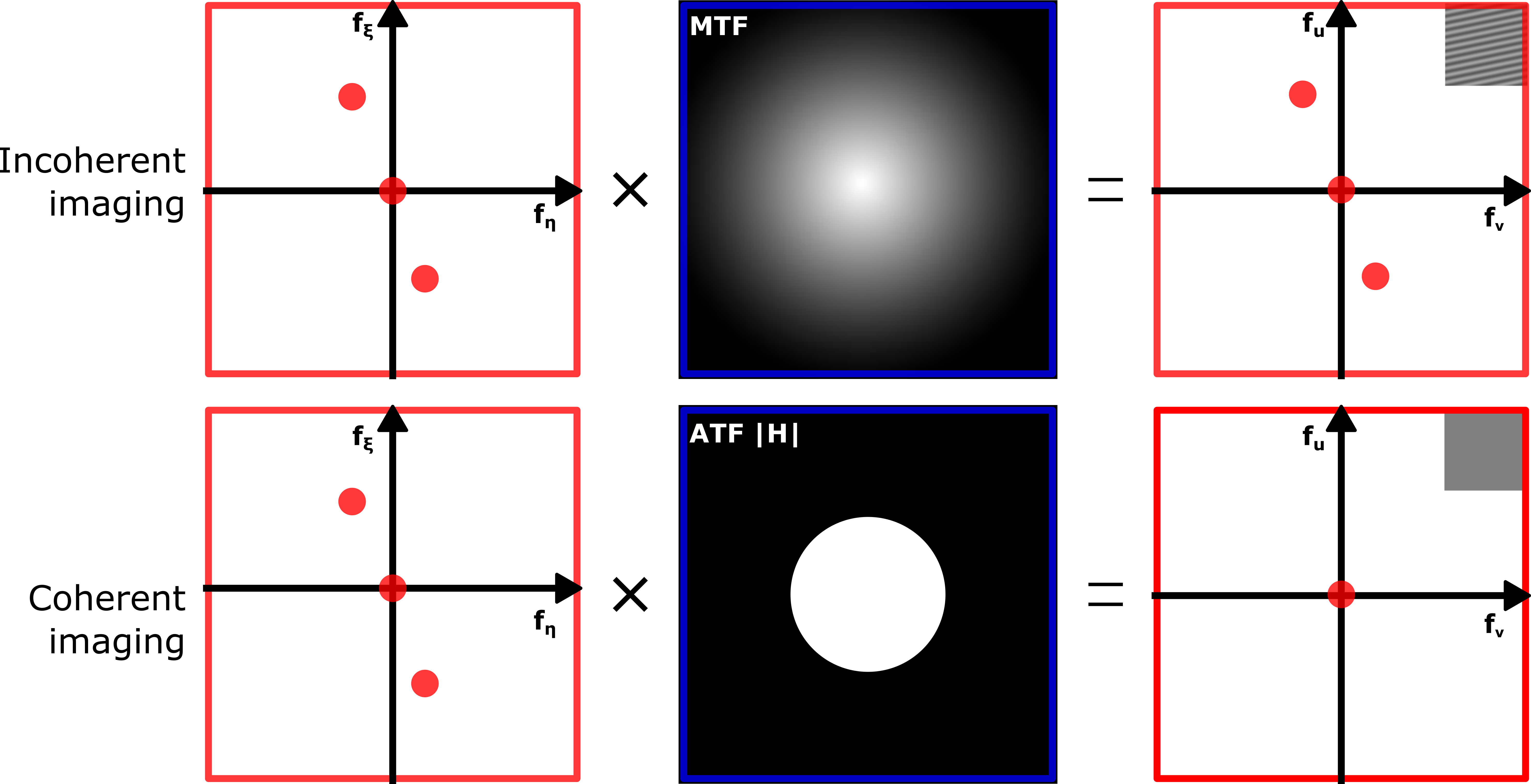}
\caption{Image formation of a high frequency modulation pattern using the concept of Fourier optics in frequency domain. The illumination is incoherent (top of the figure) and coherent (bottom of the figure) and the imaging system, aberration-free. The incoherent imaging system resolves the sinusoidal object pattern but the contrast of the resulting image is reduced by the modulation transfer function, whereas the coherent imaging system is not able to transmit the features of the object pattern.}
\label{2dcoherence}
\end{figure}
Some figures were extracted from the Kurt Thorn's presentation (\href{http://nic.ucsf.edu/dokuwiki/doku.php?id=presentations}{Microscopy optics II}" [Online: accessed 2018]).
\end{appendices}
\end{document}